\documentstyle[12pt]{article}
\begin{document}
\begin{center}

\large{{\bf Detecting $\nu_{\tau}$ Oscillations at $PeV$ Energies}}\\

\bigskip

John~G.~Learned and Sandip~Pakvasa\\

{\it Department of Physics and Astronomy, Univerity of Hawaii}\\
\bigskip
\small{\it{Preprint DUMAND-3-94}}\\

\small{\it{15 May 1994}}\\

\bigskip

Abstract
\end{center}

It is suggested that a large deep underocean neutrino detector, given the
presence of significant numbers of neutrinos in the PeV range as predicted by
various models of Active Galactic Nuclei, can make unique measurements of the
properties of neutrinos.  It will be possible to observe the existence of the
$\nu_{\tau}$, measure its mixing with other flavors, in fact test the mixing
pattern for all three flavors based upon the mixing parameters suggested by the
atmospheric and solar neutrino data, and measure the $nu_{\tau}$ cross section.
The key signature is the charged current $\nu_{\tau}$ interaction, which
produces a double cascade, one at either end of a minimum ionizing track.  At a
few PeV these cascades would be separated by roughly $100\ m$, and thus be
easily resolvable in DUMAND and similar detectors.  Future applications are
precise neutrino astronomy and earth tomography.

\section{The Double Bang Signal}

\begin{figure}
\setlength{\unitlength}{0.012500in}%
\begingroup\makeatletter\ifx\SetFigFont\undefined
\def\x#1#2#3#4#5#6#7\relax{\def\x{#1#2#3#4#5#6}}%
\expandafter\x\fmtname xxxxxx\relax \def\y{splain}%
\ifx\x\y   
\gdef\SetFigFont#1#2#3{%
  \ifnum #1<17\tiny\else \ifnum #1<20\small\else
  \ifnum #1<24\normalsize\else \ifnum #1<29\large\else
  \ifnum #1<34\Large\else \ifnum #1<41\LARGE\else
     \huge\fi\fi\fi\fi\fi\fi
  \csname #3\endcsname}%
\else
\gdef\SetFigFont#1#2#3{\begingroup
  \count@#1\relax \ifnum 25<\count@\count@25\fi
  \def\x{\endgroup\@setsize\SetFigFont{#2pt}}%
  \expandafter\x
    \csname \romannumeral\the\count@ pt\expandafter\endcsname
    \csname @\romannumeral\the\count@ pt\endcsname
  \csname #3\endcsname}%
\fi
\fi\endgroup
\begin{picture}(422,325)(78,440)
\thinlines
\put(141,699){\circle*{6}}
\put(141,657){\circle*{6}}
\put(141,614){\circle*{6}}
\put(141,572){\circle*{6}}
\put(141,530){\circle*{6}}
\put(141,488){\circle*{6}}
\put(141,445){\circle*{6}}
\put(310,741){\circle*{6}}
\put(310,699){\circle*{6}}
\put(310,657){\circle*{6}}
\put(310,614){\circle*{6}}
\put(310,572){\circle*{6}}
\put(310,530){\circle*{6}}
\put(310,488){\circle*{6}}
\put(310,445){\circle*{6}}
\put(479,741){\circle*{6}}
\put(479,699){\circle*{6}}
\put(479,657){\circle*{6}}
\put(479,614){\circle*{6}}
\put(479,572){\circle*{6}}
\put(479,530){\circle*{6}}
\put(479,488){\circle*{6}}
\put(479,445){\circle*{6}}
\put(141,572){\circle{22}}
\put(141,614){\circle{10}}
\put(141,530){\circle{32}}
\put(141,488){\circle{22}}
\put(141,445){\circle{10}}
\put(310,614){\circle{10}}
\put(141,741){\circle*{6}}
\put(310,657){\circle{10}}
\put(109,464){\makebox(0,0)[lb]{\smash{\SetFigFont{12}{14.4}{rm}Neutrino
Interaction}}}
\put(310,572){\circle{12}}
\put(141,657){\circle{10}}
\put(479,699){\circle{42}}
\put(479,741){\circle{32}}
\put(479,657){\circle{32}}
\put(479,614){\circle{22}}
\put(479,572){\circle{12}}
\thicklines
\put( 78,509){\line( 2, 1){392}}
\thinlines
\put(141,762){\line( 0,-1){317}}
\put(310,765){\line( 0,-1){320}}
\put(479,762){\line( 0,-1){317}}
\put(143,607){\vector( 1, 2){0}}
\multiput(
99,519)(3.98182,7.96364){11}{\makebox(0.1111,0.7778){\SetFigFont{5}{6}{rm}.}}
\put(138,567){\vector( 1, 2){0}}
\multiput(120,530)(3.68000,7.36000){5}{\makebox(0.1111,0.7778){\SetFigFont{5}{6}{rm}.}}
\put(138,530){\vector( 4,-1){0}}
\multiput(130,532)(8.00000,-2.00000){1}{\makebox(0.1111,0.7778){\SetFigFont{5}{6}{rm}.}}
\put(136,490){\vector( 3,-4){0}}
\multiput(110,525)(5.23200,-6.97600){5}{\makebox(0.1111,0.7778){\SetFigFont{5}{6}{rm}.}}
\put(306,652){\vector( 2, 3){0}}
\multiput(275,605)(5.17948,7.76922){6}{\makebox(0.1111,0.7778){\SetFigFont{5}{6}{rm}.}}
\put(305,611){\vector( 4,-1){0}}
\multiput(290,615)(7.41175,-1.85294){2}{\makebox(0.1111,0.7778){\SetFigFont{5}{6}{rm}.}}
\put(305,575){\vector( 4,-1){0}}
\multiput(245,590)(8.57143,-2.14286){7}{\makebox(0.1111,0.7778){\SetFigFont{5}{6}{rm}.}}
\put(479,738){\vector( 1, 2){0}}
\multiput(460,700)(3.80000,7.60000){5}{\makebox(0.1111,0.7778){\SetFigFont{5}{6}{rm}.}}
\put(478,662){\vector( 1,-1){0}}
\multiput(445,695)(6.50000,-6.50000){5}{\makebox(0.1111,0.7778){\SetFigFont{5}{6}{rm}.}}
\put(475,619){\vector( 1,-2){0}}
\multiput(440,690)(3.93333,-7.86667){9}{\makebox(0.1111,0.7778){\SetFigFont{5}{6}{rm}.}}
\put(101,469){\line(-1, 0){ 15}}
\put( 86,469){\vector( 1, 3){ 14.100}}
\put(197,538){\line(-1, 0){ 11}}
\put(186,538){\vector( 1, 2){  8.800}}
\put(411,725){\line( 1, 0){ 18}}
\put(429,725){\vector( 1,-1){ 21}}
\thicklines
\put( 80,510){\line( 6, 5){ 30}}
\put(110,535){\line( 1, 0){  5}}
\multiput(115,535)(0.25000,-0.50000){21}{\makebox(0.4444,0.6667){\SetFigFont{7}{8.4}{rm}.}}
\put(120,525){\line( 0,-1){  5}}
\put(120,520){\line(-4,-1){ 40}}
\put( 80,510){\line( 0, 1){  0}}
\put(430,685){\line( 6, 5){ 30}}
\put(460,710){\line( 1, 0){  5}}
\multiput(465,710)(0.25000,-0.50000){21}{\makebox(0.4444,0.6667){\SetFigFont{7}{8.4}{rm}.}}
\put(470,700){\line( 0,-1){  5}}
\put(470,695){\line(-4,-1){ 40}}
\put(430,685){\line( 0, 1){  0}}
\put(344,720){\makebox(0,0)[lb]{\smash{\SetFigFont{12}{14.4}{rm}Tau Decay}}}
\put(204,532){\makebox(0,0)[lb]{\smash{\SetFigFont{12}{14.4}{rm}Tau Track}}}
\end{picture}

\caption{A schematic view of a ``double bang'' event near a deep ocean
detector whose modules are indicated by dots.}
\label{fig:DOUBLEBANG}
\end{figure}
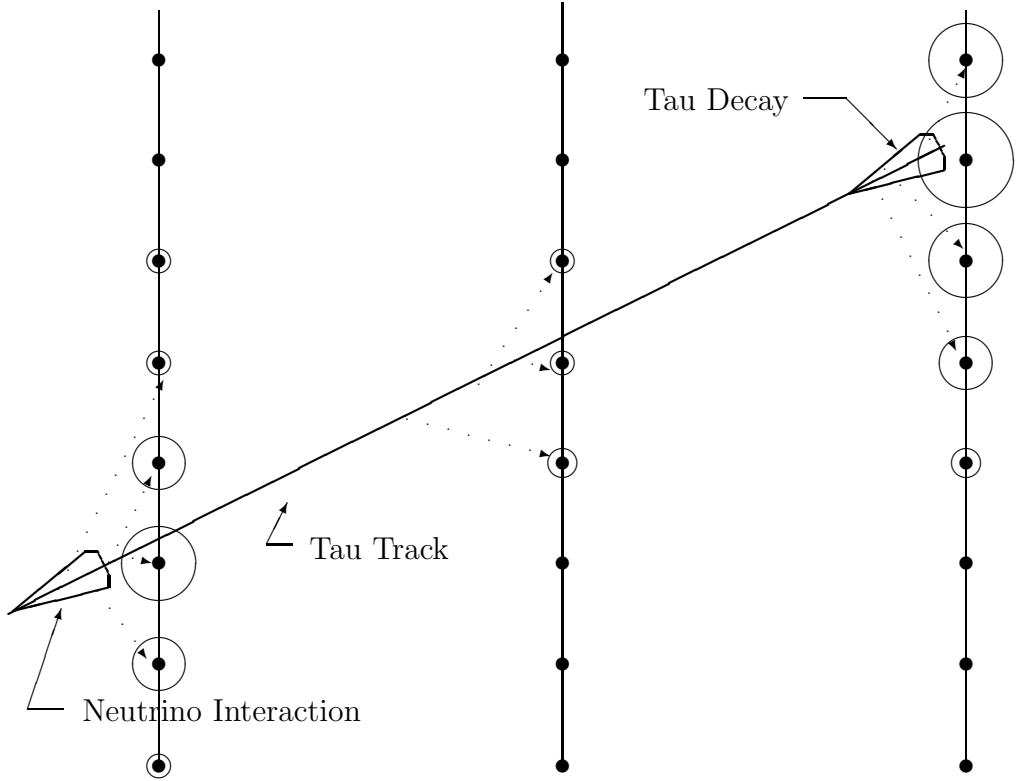

In thinking about the consequences of $\tau$ production in DUMAND it has become
clear that we may find a spectacular signature for $\tau$ events.  This depends
upon the existence of $10^{15}\ eV$ neutrinos in  adequate numbers, as are in
fact predicted from AGNs\cite{STENGER92} for example.  Since the $\tau$ mass is
about $1.8\ GeV$, a $\tau$ of $1.8\ PeV$ and with $c \tau$ of  $91\ \mu
m$\cite{BIBLE92} would fly roughly $90\ m$ before self destructing.  The
interesting signals are from the charged current quark interactions of
$\nu_{\tau}$'s.  The signature, as illustrated in Figure \ref{fig:DOUBLEBANG},
is:

\begin{enumerate}

\item    a big hadronic shower from the initial $\nu_{\tau}$ interaction,

\item    a muon like $\tau$ track,  and then

\item    a second big particle cascade (usually 3 times larger).

\end{enumerate}

To give some scale to this, the ratios of detectable photons from these three
segments are roughly $10^{12}$ : $2\cdot 10^6$ : $3\cdot 10^{11}$.

The charged $\tau$ will be hard to resolve from the bright light and not very
different times from the cascades, but simply connecting the two cascades by
the speed of light will suffice to make an unambiguous association.  This
appears to be a unique signature for real $\tau$ production by $\nu_{\tau}$'s,
thus ``discovering'' the $\nu_{\tau}$, and infering the mixing angles for
neutrino oscillations. Finding even one of these events would have significant
implications.

The point is that in the general energy range of a few $PeV$ there exists a
powerful tool for searching for $\tau$ mixing, over an unequalled parameter
space, with unambiguous identification of the $\tau$. We know of no other way
to make a $\nu_{\tau}$ {\bf appearance} experiment with the cosmic rays, no way
has been proposed for an accelerator experiment except for the use of emulsions
making observations of relatively large $\delta m^2$, and no way of detecting
$\nu_{\tau}$'s except statistically at proposed long baseline accelerator
experiments. (Indeed the costs of these endeavors are similar if one
includes the preparation of the beams at the accelerators.)

In the following we explore the physics implications of the observations of the
double bang events in a little more detail, discussing the kinematics, the
sensitivity to two and three neutrino mixing, and potential backgrounds.

\section{Essentially Full Kinematics from Double Bang Events}

One gets to measure the total energy of the incident neutrino and nearly the
full kinematics of the double bang events. The first cascade yields the energy
transfered to the quark, and the second cascade gives the energy kept by the
$\tau$; the sum give the total incoming neutrino energy, and the ratio of the
first cascade energy to the total energy provides the $y$ value.  The cross
sections and $<y>$ are almost equal for $\nu_{\tau}$ and
$\bar{\nu}_{\tau}$ at this energy\cite{QUIGG86}.  Observing the $y$
distribution is a check on the observations, and departures from expectations
could signal new physics. In calculating the $\nu_{\tau}$ flux, the measured
$y$ distribution will permit correction for the potentially unobserved events
near $y\ =\ 0$ (no initial cascade) and near $y\ =\ 1$ (initial cascade most of
the energy and the tau decays too close for resolution).  The near equality of
the cross sections for particle and antiparticle permits the total flux to be
calculated independently of the mix in the cosmic beam.

The threshold energy for discriminating two cascades will be determined by
having a $\tau$ that flies far enough so that the two cascades can be
distinguished, and so that there are no ``punch through'' events.  This
distance will be of the order of some few times the cascade length (order $10\
m$), and thus our threshold for $\tau$ detection via this means would be, as
suggested above, about a $PeV$.   With the physics limitation of several tens
of meters, there will also be a detector dependent limitation depending upon
detector density and response.

We note that the observation of the double bang events presents the opportunity
to measure the $PeV$ $\nu_{\tau}$ cross section via the angular distribution in
the lower hemispere caused by attenuation through the earth ($\sim 90\%$ in
this energy range).  For future studies of earth tomography, the potential of
this process is great, since it does not depend upon convolution over the $y$
distribution and muon range, as is necessary to extract informatiom from the
upcoming muon flux alone.

Also, given the enormous light output of the cascades one would expect that
timing from the detectors (at intermediate distances, since nearby detectors of
present design will surely be saturated) would give excellent vertex
resolution, and thus the initial neutrino direction to a precision of order on
$1^{\circ}$.  In principle, of course, having both cascades and almost all
energy ``visible'' one can deduce the initial neutrino direction with arbitrary
precision, perhaps making optical precision ultimately possible in some future
neutrino telescope.

\section{Deducing the Neutrino Flux Flavor Content}

{}From the ensemble of measurements with a DUMAND like array we will have:

\begin{enumerate}

\item 	The $\tau$ rate from double bang events gives the $\nu_{\tau} +
\bar{\nu}_{\tau}$ flux.

\item 	Measuring the UHE muon flux permits calculating the $\nu_{\mu} +
\bar{\nu}_{\mu}$ flux.

\item 	The $W^-$ resonant event rate yields the $\bar{\nu}_e$ flux at
$6.4\ PeV$.

\item 	Measuring the cascade rate (as a function of energy) gives the sum of
neutral current interactions of all flavors of neutrinos and charged currents
without visible $\mu$'s and $\tau$'s, that is mostly $\nu_e$'s.

\end{enumerate}

If we write $r\ \equiv\ \sigma_{CC} / \sigma_{NC}$, the ratio of charged to
neutral current cross sections, and we note that the cross sections are nearly
flavor and charge independent at this energy, we can summarize the four
observations abov as:

\begin{equation}
N_1 = N_{\tau} + N_{\bar{\tau}}
\end{equation}

\begin{equation}
N_2 = N_{\mu} + N_{\bar{\mu}}
\end{equation}

\begin{equation}
N_3 = N_{\bar{e}}
\end{equation}

\begin{equation}
N_4 = (N_{e} + N_{\bar{e}})\cdot (1 + r) +
(N_{\mu} + N_{\bar{\mu}}) +
(N_{\tau} + N_{\bar{\tau}})
\end{equation}

Combining these four equations we can then extract 3 flavor fractions
($f_e\      =\ (N_e+N_{\bar{e}})/N_{total}$,
$f_{\mu}\  =\ (N_{\mu}+N_{\bar{\mu}})/N_{total}$,  and
$f_{\tau}\ =\ (N_{\tau}+N_{\bar{\tau}})/N_{total}$),
and the antiparticle to particle ratio, $N_{\bar{e}}/N_e$.

We expect nearly equal numbers of particles and anti-particles, with half as
many $\nu_e$'s as $\nu_{\mu}$'s.  In some scenarios there would be few
$\nu_e$'s.   The $\nu_{\tau}$ flux from any source is expected to be very small
in any case. Most predicted scenarios indicate adequate flight time for the
$\pi$'s and the $\mu$'s to decay at the source.  In that circumstance there is
only one free number in the initial flux ratios, the initial  $\pi^+$ to
$\pi^-$  ratio.

The quantities to be observed do have different energy dependent sensitivity,
which will complicate matters.  We would expect that systematic uncertainties
(in factors such as the effective volume, energy calibration, etc.) will be
important in interpreting results.  Detector specific simulations are needed to
make explicit evaluations of these measurement possibilities.  Nonetheless, it
seems to us to be practical to extract the total flavor content of the cosmic
flux with reasonable precision from this suite of observations.

\section{Sensitivity to Neutrino Oscillations}

Our $\delta m^2$ sensitivity (from $L/E$) is then fantastic, going down to the
order of $10^{-16}\ eV^2$ (the distance is out to the AGNs, $\sim 100\ MPc$).
To determine the two neutrino mixing angle sensitivity limit requires detector
specific simulations.  The limitation has to do with the AGN neutrino flux
magnitude and effective volume for these events, and will probably be limited
by statistics, at least in the near future.  We guess it will be no better than
$0.01$ in the best of situations (as with most experiments).

To discuss the (to be observed) fluxes in terms of neutrino oscillations we
make a number of simplifying (but reasonable) assumptions.  Explicitly, we
assume that:

\begin{enumerate}

\item the initial fluxes are in the proportion
$\nu_{\mu}$:$\nu_{e}$:$\nu_{\tau}$::2:1:0 (as generally expected),

\item there are equal numbers of neutrinos and anti-neutrinos (although this is
not crucial and can easily be dropped in actual analysis),

\item all the $\delta m^2$ are above $10^{-16}\ eV^2$, and that
$sin^2(\delta m^2 L/4 E)$ all average to $1/2$,

\item matter effects at the production site are negligible (this is reasonable
since $N_{e^-} \simeq N_{e^+}$ and baryon densities are low), and

\item  that there are no large matter effects in the path to earth.  The latter
is reasonable in the $\delta m^2$ range of interest ($10^{-2}$ to
$10^{-6}\ eV^2$).

\end{enumerate}

With these assumptions, let us first consider the case where the low energy
atmospheric neutrino anomaly is accounted for by neutrino oscillations.  The
oscillations could be either $\nu_{\mu} \leftrightarrow \nu_{e}$ or
$\nu_{\mu} \leftrightarrow \nu_{\tau}$ with $\delta m^2$ of $10^{-2}-10^{-3}\
eV^2$ and $sin^2
2 \theta$ ranging from $0.5$ to $1.0$.

In both cases the $\nu_{\mu}/\nu_e$ ratio should be modified {\bf exactly} as
in the atmospheric case:  given that the $\nu_{\mu}/\nu_e$ is found to be $0.6$
of the expected value of $2$, we expect exactly the same thing from the distant
cosmic sources of much higher energy, $\nu_{\mu}$:$\nu_e$ = 1.2:1. In the
$\nu_{\mu} \leftrightarrow \nu_e$ mixing case this translates into
$\nu_{\mu}$:$\nu_{e}$:$\nu_{\tau}$::1.64:1.36:0 (using the earlier
normalization to 3 total). In the other case, of $\nu_{\mu} \leftrightarrow
\nu_{\tau}$ mixing, we predict  $\nu_{\mu}$:$\nu_{e}$:$\nu_{\tau}$::1.2:1:0.8.
It is easy to see that in the two flavor mixing case, $\nu_e/\nu_{\mu}$ can
never be greater than one.  Even when the mixing is maximal on finds
$\nu_{\mu}$:$\nu_{e}$:$\nu_{\tau}$::1.5:1.5:0 in the $\nu_{\mu} \leftrightarrow
\nu_e$ mixing case, and $\nu_{\mu}$:$\nu_{e}$:$\nu_{\tau}$::1:1:1 in the
$\nu_{\mu} \leftrightarrow \nu_{\tau}$ case.

We can now ask how expectations will change if we include neutrino oscillation
solutions to the solar neutrino deficit.  We can consider two distinct
possibilities:

\begin{enumerate}

\item If the $\nu_e$ mixing with is with with either other species with a small
angle, as in the small angle MSW regime, then the above results remain
unaffected.

\item If all three flavors mix substantially and the $\delta m^2$'s are in the
range of $10^{-5}$ to $10^{-6}\ eV^2$ (corresponding to the MSW ``large angle''
solution) or $10^{-10}\ eV^2$ (corresponding to the ``long wavelength'' case).

\end{enumerate}

\begin{figure}
\setlength{\unitlength}{0.012500in}%
\begingroup\makeatletter\ifx\SetFigFont\undefined
\def\x#1#2#3#4#5#6#7\relax{\def\x{#1#2#3#4#5#6}}%
\expandafter\x\fmtname xxxxxx\relax \def\y{splain}%
\ifx\x\y   
\gdef\SetFigFont#1#2#3{%
  \ifnum #1<17\tiny\else \ifnum #1<20\small\else
  \ifnum #1<24\normalsize\else \ifnum #1<29\large\else
  \ifnum #1<34\Large\else \ifnum #1<41\LARGE\else
     \huge\fi\fi\fi\fi\fi\fi
  \csname #3\endcsname}%
\else
\gdef\SetFigFont#1#2#3{\begingroup
  \count@#1\relax \ifnum 25<\count@\count@25\fi
  \def\x{\endgroup\@setsize\SetFigFont{#2pt}}%
  \expandafter\x
    \csname \romannumeral\the\count@ pt\expandafter\endcsname
    \csname @\romannumeral\the\count@ pt\endcsname
  \csname #3\endcsname}%
\fi
\fi\endgroup
\begin{picture}(370,376)(8,447)
\thinlines
\put(271,613){\circle{8}}
\put(219,665){\circle{8}}
\put( 61,823){\line( 1,-1){317}}
\put( 61,506){\line( 1, 1){158.500}}
\put( 61,669){\line( 1,-1){160.500}}
\thicklines
\put(168,613){\line( 1, 1){ 51.500}}
\put( 61,823){\line( 0,-1){317}}
\put( 61,506){\line( 1, 0){317}}
\thinlines
\put(251,696){\vector(-1,-1){ 19.500}}
\put(287,641){\vector(-2,-3){ 12.923}}
\put(176,657){\vector(-1,-1){ 16}}
\put(211,538){\vector(-1, 2){ 15.800}}
\put(184,649){\line(-1,-1){ 24}}
\put(160,625){\line( 0,-1){ 16}}
\put(160,609){\line( 1,-1){ 27.500}}
\put(188,582){\line( 3,-1){ 83.100}}
\put(271,554){\line( 0, 1){ 20}}
\put(271,574){\line(-5, 4){ 43.902}}
\put(227,609){\line(-3, 1){ 38.700}}
\put(188,621){\line( 2, 3){ 10.769}}
\put(199,637){\line(-5, 4){ 15}}
\put(184,649){\line( 0, 1){  0}}
\put(188,613){\line(-1, 1){ 20}}
\multiput(168,633)(-0.40000,-0.40000){21}{\makebox(0.1111,0.7778){\SetFigFont{5}{6}{rm}.}}
\put(160,625){\line( 0,-1){ 16}}
\put(160,609){\line( 4, 3){ 16}}
\multiput(176,621)(0.48000,-0.32000){26}{\makebox(0.1111,0.7778){\SetFigFont{5}{6}{rm}.}}
\put(188,613){\line( 0, 1){  0}}
\put(156,447){\makebox(0,0)[lb]{\smash{\SetFigFont{14}{16.8}{rm}Muon
Fraction}}}
\put(168,613){\circle{8}}
\put( 25,601){\makebox(0,0)[lb]{\smash{
\SetFigFont{14}{16.8}{rm}Electron
Fraction
}}}
\put(219,517){\makebox(0,0)[lb]{\smash{\SetFigFont{14}{16.8}{rm}Allowed
Region}}}
\put(295,657){\makebox(0,0)[lb]{\smash{\SetFigFont{14}{16.8}{rm}Expected}}}
\put(295,637){\makebox(0,0)[lb]{\smash{\SetFigFont{14}{16.8}{rm}Cosmic}}}
\put( 57,479){\makebox(0,0)[lb]{\smash{\SetFigFont{14}{16.8}{rm}0.0}}}
\put(370,479){\makebox(0,0)[lb]{\smash{\SetFigFont{14}{16.8}{rm}1.0}}}
\put( 49,502){\makebox(0,0)[lb]{\smash{
\SetFigFont{14}{16.8}{rm}0.0
}}}
\put( 53,815){\makebox(0,0)[lb]{\smash{
\SetFigFont{14}{16.8}{rm}1.0
}}}
\put(207,665){\makebox(0,0)[lb]{\smash{
\SetFigFont{14}{16.8}{rm}0.0
}}}
\put(128,586){\makebox(0,0)[lb]{\smash{
\SetFigFont{14}{16.8}{rm}0.5
}}}
\put(132,625){\makebox(0,0)[lb]{\smash{
\SetFigFont{14}{16.8}{rm}Tau
Fraction
}}}
\put(259,716){\makebox(0,0)[lb]{\smash{\SetFigFont{14}{16.8}{rm}Observed }}}
\put(259,696){\makebox(0,0)[lb]{\smash{\SetFigFont{14}{16.8}{rm}Underground}}}
\put(295,617){\makebox(0,0)[lb]{\smash{\SetFigFont{14}{16.8}{rm}no Osc}}}
\put(219,538){\makebox(0,0)[lb]{\smash{\SetFigFont{14}{16.8}{rm}3 Neutrino
Mix}}}
\end{picture}

\caption{The fraction of muon neutrinos versus electron neutrinos, allowing for
a fraction of tau neutrinos.  The expected initial flux is at 0.66, 0.33.  Full
and equal mixing would result in 0.33, 0.33 (and 0.33 taus).  The points
represent calculations of results for various solutions to the solar and
atmospheric neutrino problems.}
\label{fig:FRACTIONS}
\end{figure}
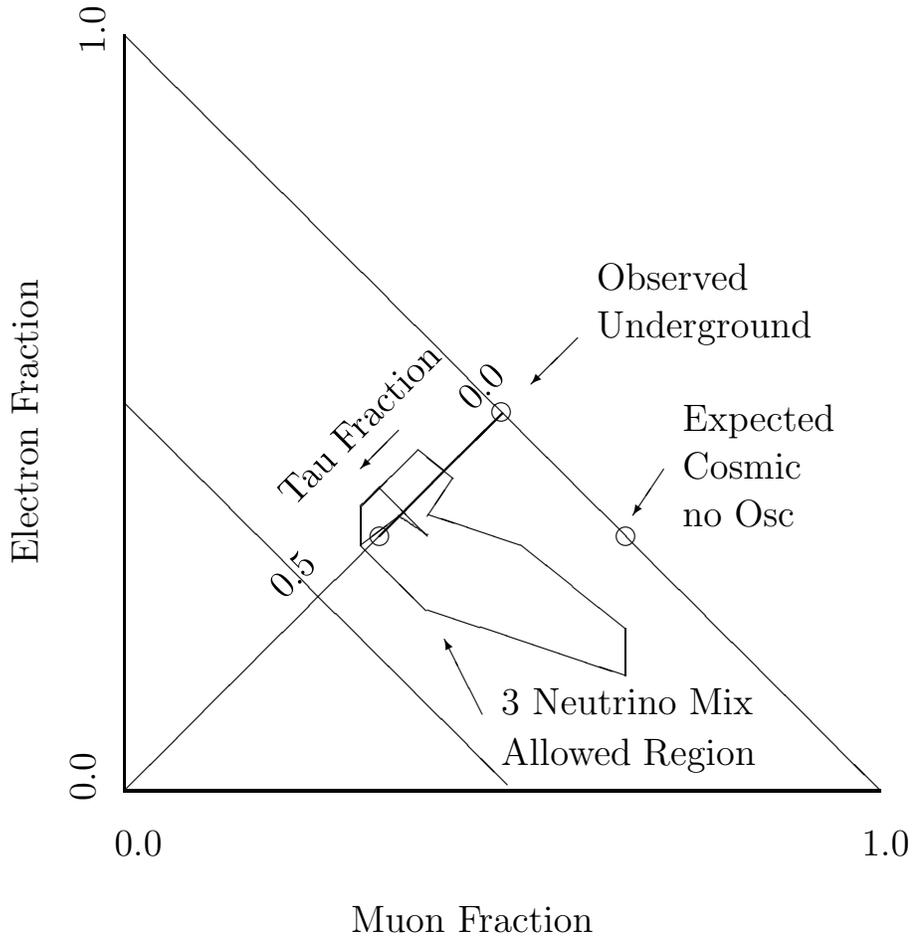

The parameter space for a combined fit to the atmospheric and solar neutrino
data has been given for each of the above cases by  Fogli, {\it et
al.}\cite{FOGLI93}, and by Acker, {\it et al.}\cite{ACKER93}, respectively.  We
have evaluated the survival and transition probabilities for the whole range of
allowed values of the three mixing angles, and the results are shown in Figure
\ref{fig:FRACTIONS}.  This figure plots the fraction of muon neutrinos versus
the fraction of electron neutrinos, and thus where each point specifies a
fraction of tau neutrinos  (it is analogous to the color triangle). The initial
expected flux ($\nu_{\mu}$:$\nu_{e}$:$\nu_{\tau}$::2:1:0) is at $f_{\mu}\ =\
0.66$ and $f_e\ =\ 0.33$.

We further observe that:

\begin{enumerate}

\item Amost all combinations of acceptable mixing angles result in saturation
values to be observed with AGN neutrinos which lie between the lines
$f_e+f_{\mu}\ =\ 0.88$ and  $f_e+f_{\mu}\ =\ 0.66$.  Hence a substantial
number of $\nu_{\tau}$ events are expected ($0.34 > f_{\tau} > 0.12$) for all
situations which solve the solar and atmospheric problems with neutrino
oscillations.

\item Since in the case of two flavor mixing it is impossible to obtain
$f_e/f_{\mu} > 1$, observation of the data falling above the diagonal
$f_e = f_{\mu}$ line is clear evidence for three flavor mixing.

\end{enumerate}

\section{Backgrounds: Almost None}

We do not know of any particle with the ability to penetrate $\sim 10^4\
gm/cm^2$ of matter before decaying or interacting  ($100\ m$ is about $100$
strong interaction lengths in water), except for  leptons.  Of the leptons,
neutrinos have interaction lengths so great as to be negligibly likely to
interact at close range ($<10^{-6}$ at this $PeV$ energy).  Electrons will
immediately radiate.  Muons are the most likely to cause confusion.  A muon of
$2\ PeV$ energy has a mean energy loss rate of $500\ GeV/m$, and an effective
radiation length of about $1\ m$.  Since on the order of $1/2$ of the energy
loss to the medium is continuous on the scale of a meter (it is a fuzzy number
because of the continuous distribution of energy transfer fractions), the muon
track light output will be roughly $1000$ times minimum. Thus, even
though muon radiation fluctuates greatly, we can anticipate that confusion with
tau decays will be small.  Once again, detector dependent simulations are
required to numerically evaluate the confusion probability.  Whatever that
probability is, and we expect it to be low at $2\ PeV$, it decreases rapidly
with energy (short range muon radiation goes up, and tau decay length
increases).  Finally, studies of the path length distribution can confirm the
observation of tau decays, requiring consistency with the known tau lifetime.

An other potential background might be due to downgoing muons. The downgoing
muons are strongly peaked near the zenith, while the double bang events should
be uniform in direction in the upper hemisphere, but heavily depleted from
below the horizon by earth attenuation.  A stronger constraint is that events
with this energy are not expected from downgoing muons.  Bremmstrahlung events
of a few hundred $GeV$ are plentiful, but the flux falls very fast with energy,
and negligible numbers are expected above  $100\ TeV$\cite{OKADA94} (the
radiation length for this energy muon is order of $20\ m$ in water.).  Thus
even if one only had order of magnitude precision in cascade energy
reconstruction (and we should have much better for near contained events such
as we are discussing), one easily eliminates muons from the surface.

\section{Old Idea, New Combination and Impact}

The above is not entirely a new idea. Several authors have written about $\tau$
signatures in the past\cite{LEARNED80}. The new recognition is in coupling the
uniqueness of the double bang signature, the encouraging flux predictions from
AGNs, plus the modern hints about neutrino oscillations, and thus being able to
make some claim as to the region we can probe in mixing space.  Further we now
recognize that observation of this novel class of interactions makes possible
more precise measurements of the neutrino cross section and earth tomography
than have been thought possible, because one can determine the {\it neutrino}
total energy and direction with a precision limited only by the detector. For
one thing, this potential observation does give motivation to filling in the
volume, somewhat, of the hypothetical  $1\ km^3$ array, and perhaps itself
could justify construction of that experiment.  Also, given that these events
are near the acoustic detection threshold, one may contemplate hearing the
double clicks from such events at $km$ ranges and higher energies. (Bottom and
surface reflected pulses might be thought to be a confusion factor, but in fact
most often we would expect them to be useful as a tool for reconstructing range
and orientation).

We remind the reader that all of the above requires the existence of
substantial numbers of $PeV$ neutrinos, which matter should be resolved in the
next few years by the AMANDA, Baikal, DUMAND, and NESTOR experiments now under
construction. If those ultra high energy neutrinos are present in expected
numbers, then we believe that the observation of the double bang events, along
with other previously discussed interactions, will lead to important particle
physics measurements which cannot be carried out in any other way on earth.

\section*{Acknowledgement}

We want to acknowledge discussions with U.H. colleagues helping us to clarify
these matters.


\end{document}